\documentclass[twoside]{IEEEtran}
\usepackage{amsmath}
\usepackage{amssymb}
\usepackage{amsfonts}
\usepackage{graphicx}
\usepackage{tabularx}
\usepackage{subfig}
\usepackage{url}
\usepackage{xcolor}
\usepackage{soul}

\newcommand{\fakeparagraph}[1]{\vspace{2mm}\noindent\textbf{#1.}}

\begin{document}

\title{Application layer coding for IoT: benefits, limitations and implementation aspects}%
\author{Magnus~Sandell,~\IEEEmembership{Senior~Member,~IEEE,}
        Usman Raza
\thanks{The authors are with the Telecommunications Research Laboratory of Toshiba Research Europe Ltd (TREL), 32 Queen Square, Bristol BS1 4ND, UK, email: \{magnus.sandell, usman.raza\}@toshiba-trel.com. }%
}%

\maketitle

\begin{abstract}
One of the key technologies for future IoT/M2M systems are low power wide area networks, which are designed to support a massive number of low-end devices often in the unlicensed shared spectrum using random access protocols. However these usually operate without centralised control and since Automatic Repeat request and acknowledgement mechanisms are not very effective due to the strict duty cycles limits and high interference in the shared bands, many packets are lost from collisions. In this paper we analyse a recently proposed application layer coding scheme, which aims to recover lost packets by introducing redundancy in the form of a fountain code. We show how latency and decoding complexity is affected by the packet loss rate but also prove that there is a limit to what can be achieved by introducing more redundancy. The analysis is backed up by simulation results.
\end{abstract}

\section{Introduction}

Low Power Wide Area (LPWA) networks are forecasted to connect a massive number of devices in future Internet of Things (IoT)/Machine-to-Machine (M2M) networks. Traditionally, cellular technologies (2G, 3G, 4G, etc.) and short range wireless technologies (WLANs, ZigBee, Bluetooth, etc.) have been used for this purpose. However their higher cost remain prohibitive for a wider adoption for applications that require inexpensive connectivity and low-end devices to monitor our cities, industry, infrastructure and logistics. The emerging LPWA technologies such as LoRaWAN~\cite{lorawan}, SIGFOX~\cite{sigfox}, Ingenu RPMA~\cite{rpma}, NB-IoT~\cite{nbiot} and NB-FI~\cite{nbfi} are designed to provide a better coverage than the existing cellular networks at significantly lower cost and power consumption. A long range of multiple kilometres saves the LPWA technologies from the hassle of deploying very dense networks and thus avoid the exorbitant cost and the maintenance effort associated with short range wireless technologies. This also enables the end devices to connect to the network directly over a single hop, simplifying the design of the protocol stacks compared to multihop wireless technologies. Both business and technological benefits of these technologies are quickly realized by industry. A number of mobile operators, independent users and crowd-sourced start-up companies are already making strides in deploying LPWA networks across the globe. 

Motivated by this fast adoption of the LPWA technologies, many recent works studied their performance and uncovered their practical limitations in providing reliable and scalable connectivity to a massive number of devices.  It has become evident that these technologies are very prone to the intratechnology interference~\cite{Georgiou17}, cross-technology interference~\cite{Krupka16}, high frame loss~\cite{ Marcelis17}, and low capacity~\cite{ Bankov16}, clearly stressing the need for additional mechanisms to increase the reliability of the LPWA technologies. These reported problems are attributed to the combination of features that are unique to some LPWA technologies and were not present in the other long range wireless cellular technologies and thus were not studied in detail earlier. These include the use of the licence-exempt ISM bands and the random access medium control protocols (such as ALOHA), as well as the transmission duty cycle limitations dictated by the regulations on the sub-GHz ISM bands across the globe. To offer an example, LoRaWAN~\cite{lorawan} and SIGFOX~\cite{sigfox}, the two popular technologies, which use the sub-GHz ISM bands and ALOHA protocol, are subject to a 1\% duty cycle limit for all wireless devices in most sub-bands in Europe. To respect this, the base stations can neither serve a large number of end devices~\cite{Adelantado17} nor acknowledge all uplink transmissions from end devices. This means that the reliability enhancing mechanisms such as Automatic Repeat Request (ARQ) are not very effective because they require the base stations to acknowledge the uplink transmissions. Pop et al.~\cite{Pop17} show that as the LoRaWAN network scales, the base stations are frequently not able to acknowledge the successfully received uplink messages due to the duty cycle limit, leading to retransmissions from the end devices and thus collisions in  the network. Due to this reason, the ARQ based schemes can harm rather than improve overall reliability of the large networks. Marcelis et al.~\cite{Marcelis17} show that application layer coding schemes improve reliability of LoRaWAN and does not require any downlink communication. 

In this paper, we analyse the behaviour of this application layer coding dubbed as DaRe ~\cite{Marcelis17}. As LPWA technologies will handle up to millions of devices, it is of utmost importance to decode the received messages in minimal time and with low complexity. To this effect, low complexity decoding techniques are presented and shown to enable quick and efficient decoding of stream of received packets by the cloud.  As the proposed encoding and decoding techniques are not tied to any particular LPWA technology, they can be applied to a wide range of low-power networks. 


\subsection*{Our contributions}
\begin{itemize}
	\item We analyse the application layer coding scheme in \cite{Marcelis17} and show how the code rate will impact system performance. We prove that for large packet loss probabilities, it is not sufficient to reduce the code rate since this will increase the interference to a critical level.
	\item We consider latency as a metric and show how this depends on the packet loss probability.
	\item We devise a novel decoding scheme which can reduce the complexity and latency at a small price in data recovery rate.
\end{itemize}

The rest of the paper is organised as follows: a brief background of LPWA is given in Section \ref{sec:background} as well as a short overview of fountain codes. The application layer coding is presented in Section \ref{sec:app_layer} and decoding is discussed in Section \ref{sec:decoding}. Finally, conclusions are drawn in Section \ref{sec:conclusions}.

\section{Background}
\label{sec:background}
This section first describes in detail the unique peculiarities and challenges of the sub-GHz LPWA technologies that result in low reliability in the networks. We then provide a brief overview of fountain codes as a solution to improve application layer reliability of IoT applications. 

\subsection{Low Reliability of LPWA Technologies}
The landscape of LPWA technologies is crowded with multiple competing technologies deployed in the license-free ISM bands as well as the licensed bands. This paper, however, focuses only on the former that are less reliable due to the reasons discussed next.  

\fakeparagraph{Link Asymmetry} Most LPWA systems are characterized by highly asymmetric links with a dominant uplink compared to the downlink. These include SIGFOX~\cite{sigfox}, Ingenu RPMA~\cite{rpma}, and IEEE 802.15.4k~\cite{ieee802.15.4}. Some other LPWA technologies like Weightless-N~\cite{weightless} do not even provide any downlink support. The reasons for link asymmetry include use of different modulation techniques in the uplink and the downlink, minimization of listening time at end devices to reduce their energy consumption and the regional spectrum regulations.  In this scenario when the downlink and the uplink transmissions are significantly unbalanced, the reliability of the uplink transmissions cannot rely much on the downlink traffic such as acknowledgements. In fact, the techniques should be used to make the uplink transmissions more robust and resilient to packet losses in isolation with the downlink. 

\fakeparagraph{Spectrum Regulations} Spectrum regulations on the use of sub-GHz ISM band across the globe are yet another contributing factor to the link asymmetry. These regulations often limit transmission power and transmission duty cycle to efficiently share the finite radio resources among the coexisting technologies. The transmission duty cycle limit implies that the end devices and the base stations can transmit only a few messages in a day. For example, SIGFOX under the strictest settings allows an end device to transmit a maximum of 140 uplink messages but receive only 4 downlink messages from a base station. Even LoRaWAN base stations can only transmit a limited number of downlink messages, preventing them from acknowledging more than a small fraction of uplink messages as the networks scale. Again due to this reason, the downlink acknowledgements cannot be relied upon to make the uplink more reliable. In case of a base station that is operating close to its transmission duty cycle limit, acknowledgements are not guaranteed against successful receptions.  This may result in more retransmissions and more congestion in the network, leading to low reliability in the network. 

\fakeparagraph{High Packet Loss and Interference} The practical trials with the large-scale LoRaWAN, one of the prominent LPWA technologies, have shown a high frame loss even in absence of high interference ~\cite{Marcelis17}. An inevitable growth in the number of LoRaWAN devices will further increase levels of intra-technology interference. It has been shown that coverage probability will drop exponentially with network size in such networks~\cite{Georgiou17} due to collision in their use of “virtual channels”. High interference is also a by-product of using simplistic MAC protocols based on ALOHA by LoRaWAN and SIGFOX. While ALOHA simplifies the design of medium access mechanism, its uncoordinated operation results in more collisions and excessive inter-network interference. As more IoT devices will start using different wireless technologies in the sub-GHz ISM bands, cross-technology interference is bounded to increase, further limiting the network reliability. 

\begin{figure*}
\includegraphics[scale=0.66]{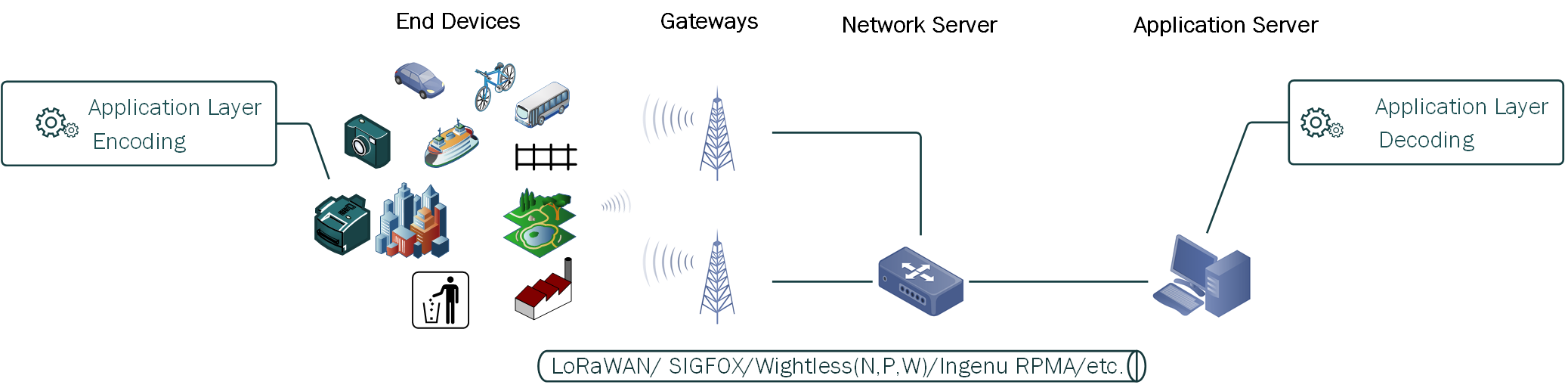}
\caption{Application layer coding over LPWA technologies}
\label{fig:archi}
\end{figure*}

Despite the limited reliability offered by many LPWA technologies, some IoT applications may still require certain reliability guarantees. In this paper, we approach this problem with a low complexity encoding/encoding technique based on fountain codes. Figure~\ref{fig:archi} shows the general architecture in which the end devices encode the application layer messages before sending them over any LPWA technology. The use of the encoding techniques on top of the LPWA technologies brings the benefit of not requiring any modification in the underlying LPWA technologies, which are often proprietary and are usually implemented in hardware. The encoded message travels through the radio access network and backend system to the application servers which decode the message. All packets are identified by a sequence number, enabling the application server to establish which packets have been lost. It can then recover these packets once a sufficient number of subsequent packets have been successfully received, so that the introduced redundancy can be exploited for decoding.

\subsection{Fountain codes\label{sec:fountain}}

A fountain code is a class of erasure codes that has the property that a (potentially) infinite sequence of encoded symbols can be generated. Conventional block erasure codes on the other hand, such as Reed-Solomon codes, have a fixed structure.  If a block of $k$ data symbols are encoded into $n$ symbols, recovery of the data is possible if any $k$ encoded symbols are received. Fountain codes have the ability to continuously generate different encoded symbols from which data recovery is possible with probability $1-2^{-k\epsilon}$ if $k(1+\epsilon)$ symbols have been received. This property makes them a $\emph{rateless code}$ since the code rate is not fixed. The small extra overhead $\epsilon$ makes it exceeding likely that the receiver can recover lost symbols and hence the receiver only needs to wait until a sufficient number of symbols are available.

Each encoded symbol is a random combination of the data symbols. Mathematically speaking, the encoded symbols $c_j$ are formed from the data symbols $d_i$ by the matrix multiplication

\begin{eqnarray}
\mathbf{c} &=& \mathbf{d} \mathbf{G} \nonumber \\
\left( c_1,\cdots,c_n \right) &=& \left( d_1,\cdots,d_k \right) \left( \mathbf{g}_1, \cdots, \mathbf{g}_k \right).
\end{eqnarray} 
The ones\footnote{In general, other Galois fields than GF(2) can be used; however in this paper we limit the discussion to binary elements.} in the vectors $\mathbf{g}_j \in \left\{0,1\right\} ^{k \times 1}$ indicate which data symbols are used create the encoded symbol $c_j$. Note that the encoding process can be done by simple Exclusive OR (XOR) operations. The received symbols will then correspond to columns of the generator matrix $\mathbf{G}$ and if $k$ linearly independent columns are available (the formed submatrix $\mathbf{G}'$ has full rank), it can be inverted and the lost data symbol are recovered. Although this can be achieved with high probability, the decoding complexity of an optimal decoder would be $\mathcal{O} \left( k^3 \right)$ making it impractical (decoding and complexity are discussed further in Section \ref{sec:decoding}). What is desired is instead linear (in $k$) encoding and decoding cost. 

The first practical fountain code achieving this was the Luby Transform (LT) code \cite{Luby02}. The decoder is based on a message passing algorithm (for more details, see Section \ref{sec:decoding}) which can find the lost symbols in linear time. To allow the decoder to process received symbols with high probability, the distribution of which data symbols are combined as encoded symbols (the ones in $\mathbf{g}_j$) is optimised. The number of data symbols, $D$, is first randomly chosen from a certain distribution and then $D$ data symbols are chosen at random. The distribution of $D$, also known as the degree distribution, was derived in \cite{Luby02}, known as the robust Soliton distribution.

Further developments on fountain codes has led to the introduction of Raptor codes. These are similar to LT codes but feature a precoding step with (usually) a conventional erasure code. The idea is that the LT code can recover a large fraction of the missing data with high probability, which then allows the outer code the recover the rest with high probability. These codes are very efficient and have been chosen for several standards, such as 3GPP MBMS \cite{3GPP-MBMS} (streaming services), DVB-H IPDC \cite{DVB-H} (IP services over DVB networks) and DVB-IPTV \cite{DVB-IPTV} (TV services over IP networks). 

\section{Application layer coding\label{sec:app_layer}}

\begin{figure}[thb]%
\centering
\includegraphics[width=\linewidth]{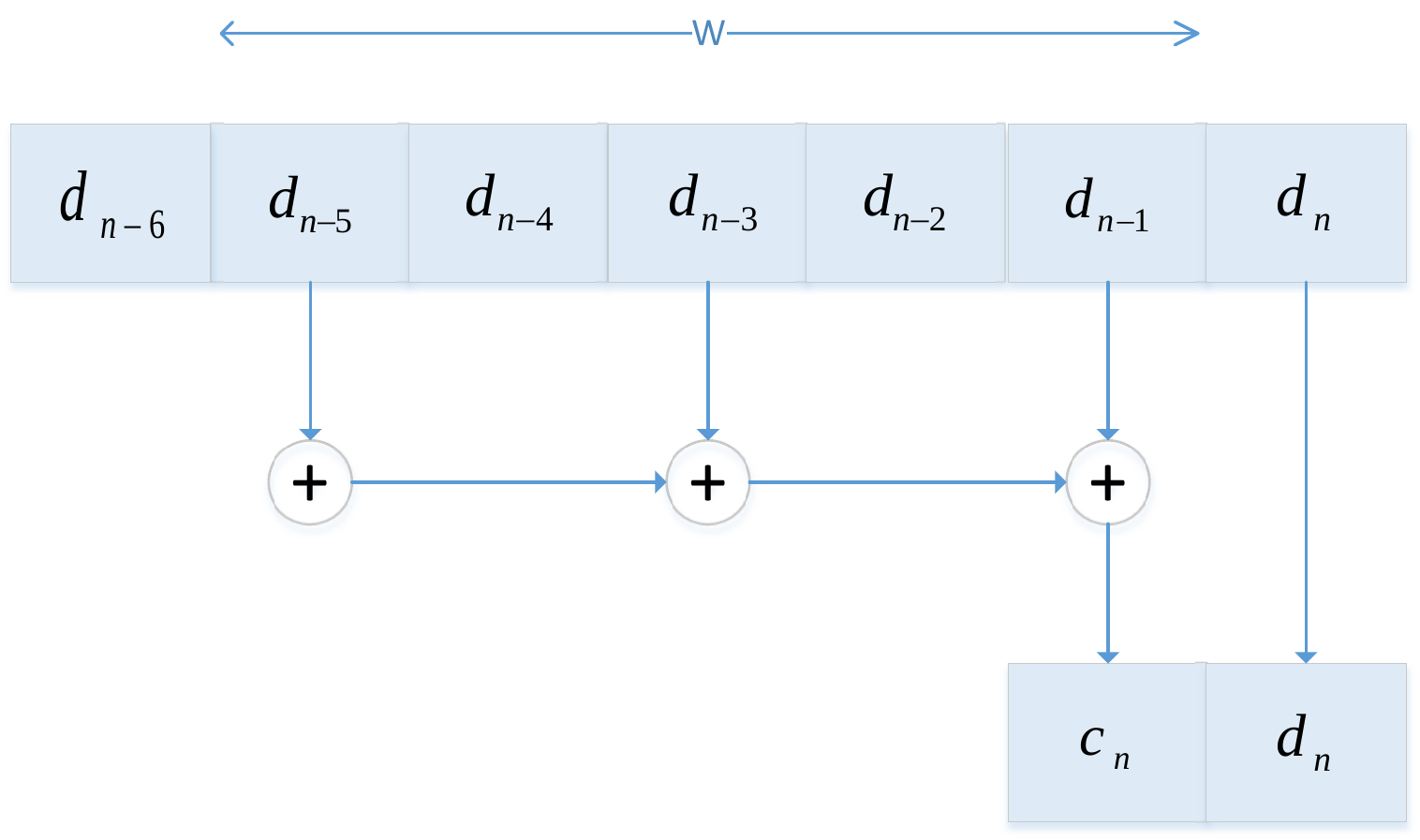}%
\caption{Data encoding principle with memory $W=5$ and degree $D=3$. The transmitted symbols are $d_n$ and $c_n$.}%
\label{fig:encoding}%
\end{figure}

To improve the performance of LoRaWAN without changing its specification, a fountain code based scheme on the application layer was proposed in \cite{Marcelis17}. In order to recover packets lost due to collisions, fading and/or shadowing, packets are amended with redundancy. This basic principle is shown in Figure \ref{fig:encoding} \cite{Marcelis17}. 

Note that we assume a systematic code, \emph{i.e.}, the data appears as one of the transmitted symbols and the redundancy is purely in the added parity symbol. Each packet carries, apart from its data, a parity symbol that is generated as a linear combination of previous data symbols. To simplify operations, all encoding is done on a bit level with XOR; this can then be repeated for all bits in the packet. Following the principle of fountain codes, the data symbols used for the redundancy changes between the different parity symbols. This random coding approach allows the receiver (on average) to receive sufficiently different linear combinations of data symbols to be able to recover the missing ones with high probability.

A few operational differences to conventional fountain codes are worth pointing out. Since the end devices continuously deliver data, there is no fixed data size $k$ since it grows with time. This mean that we need to limit the `memory' of the encoder by using a sliding window. If the encoded parity symbol is allow to depend on all previous data symbols, the end device would need to buffer all data symbols from the past. This is obviously not possible in practice, so a parity symbol $p_n$ can only depend on data symbols in a subset of $\{d_{n-W},\cdots,d_{n-1}\}$, where $W$ denotes the memory. Note that it does not make sense to make $c_n$ dependent on $d_n$ since if packet $n$ is lost, parity symbol $c_n$ is also lost and can not hence be used to recover $d_n$. One consequence of the finite memory $W$ is that the generator matrix $\mathbf{G}$ will be banded, \emph{i.e.}, only $W$ rows above the diagonal can have nonzero values. Because of this restriction, creating a degree distribution according to the LT code is difficult; in \cite{Marcelis17} a fixed degree $D$ was used. It is also worth noting that the degree distribution is designed to optimise the reduced-complexity decoder; if an optimal decoder is used, this is not necessary (for more details on the decoding, see Section \ref{sec:decoding}). Due to the finite memory $W$, the coding appears as a combination of fountain and convolutional codes. However it can also viewed as a special case of windowed erasure codes \cite{Studholme06}.  

It is worth noting that these random combinations of data symbols must be known to both the transmitter and receiver. Rather than transmitting side information for this purpose, a pseudorandom number generator can be used \mbox{\cite{Marcelis17}}. By using the same one on both the encoder and decoder side with the same seed, a synchronised pseudorandom number can be generated for each packet. This is then used to determine which data symbols make up the parity symbols.

For the example in Figure \ref{fig:encoding}, we have used memory $W=5$ and degree $D=3$ to produce one parity symbol. It is possible to extend the coding idea to produce more parity symbols which are created by different random linear combinations; with $p-1$ parity symbols the systematic would have a code rate of $R=1/p$. It is also possible to create other fractional code rate by splitting the data into $l$ segments per packet. If these are used to create $m$ parity symbols per packet, the overall code rate would be $l/(l+m)$.  

\subsection{Memory size}

\begin{figure}[thb]%
\centering
\includegraphics[width=\linewidth]{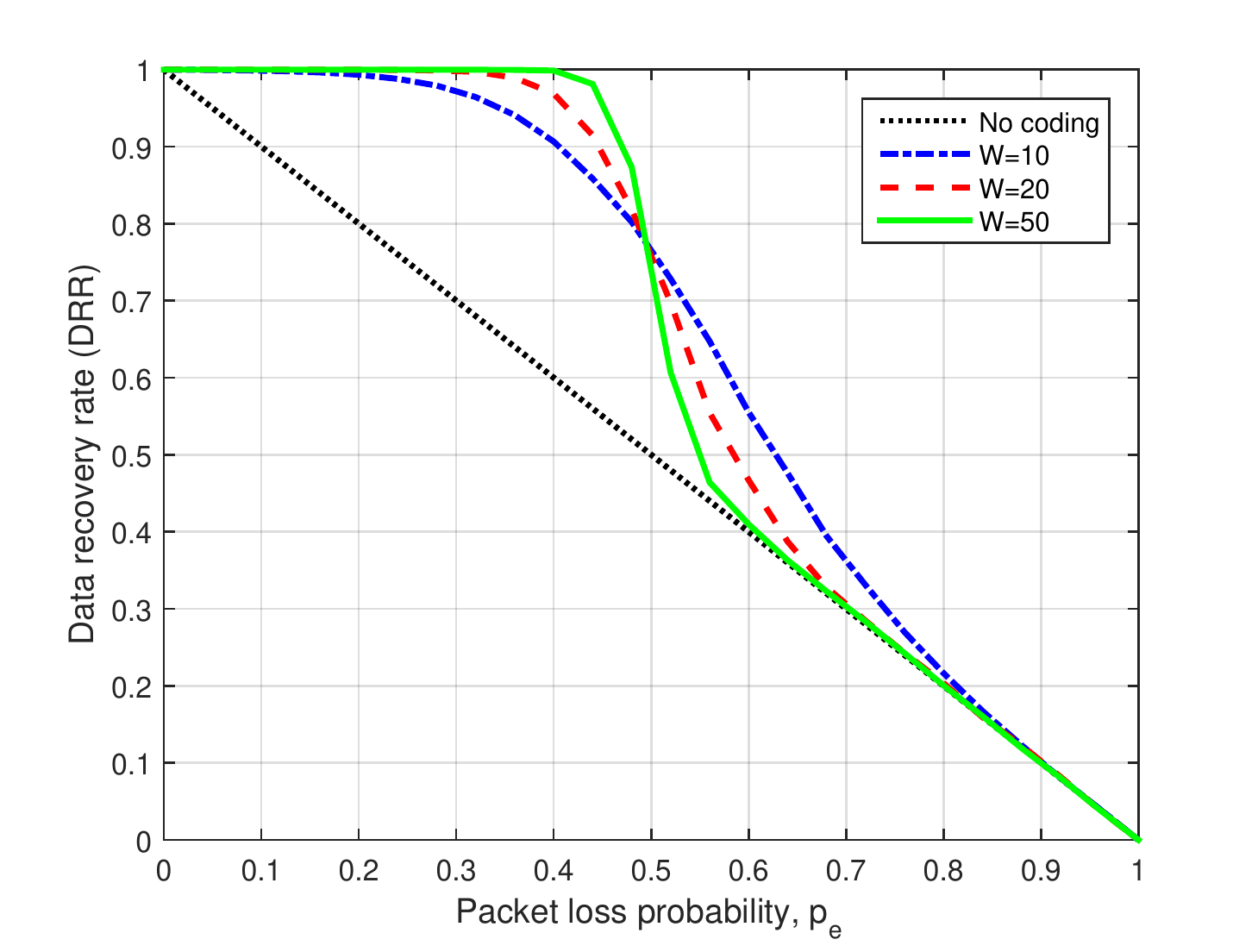}%
\caption{Data recovery rate as a function of packet loss probability ($R=1/2$, $\Delta=0.5$).}%
\label{fig:memory}%
\end{figure}

The size of the memory, $W$, will clearly have an effect on both performance and complexity. The end-devices will need to keep the last $W$ data symbols to produce the parity symbols but at the same time, larger memory offers a higher probability that a received parity symbol (column of the generator matrix) is linearly independent of the other columns. As a performance measure, we will use the Data Recovery Rate (DRR) \cite{Marcelis17} which is defined as
\begin{equation} 
\text{DRR} ~\overset{\Delta}{=}~ \frac{\text{Number of recovered data units}}{\text{Number of transmitted data units}}
\end{equation}
In Figure \ref{fig:memory} the DRR is shown for a few memory size; clearly there are diminishing returns and very little is gained by using excessively large memories. The parity density, defined as
\begin{equation}
\Delta = \frac{D}{W}
\end{equation}
is set to $\Delta=0.5$. In the next subsection, we will discuss its impact.

\subsection{Latency}

\begin{figure}[thb]%
\centering
\includegraphics[width=\linewidth]{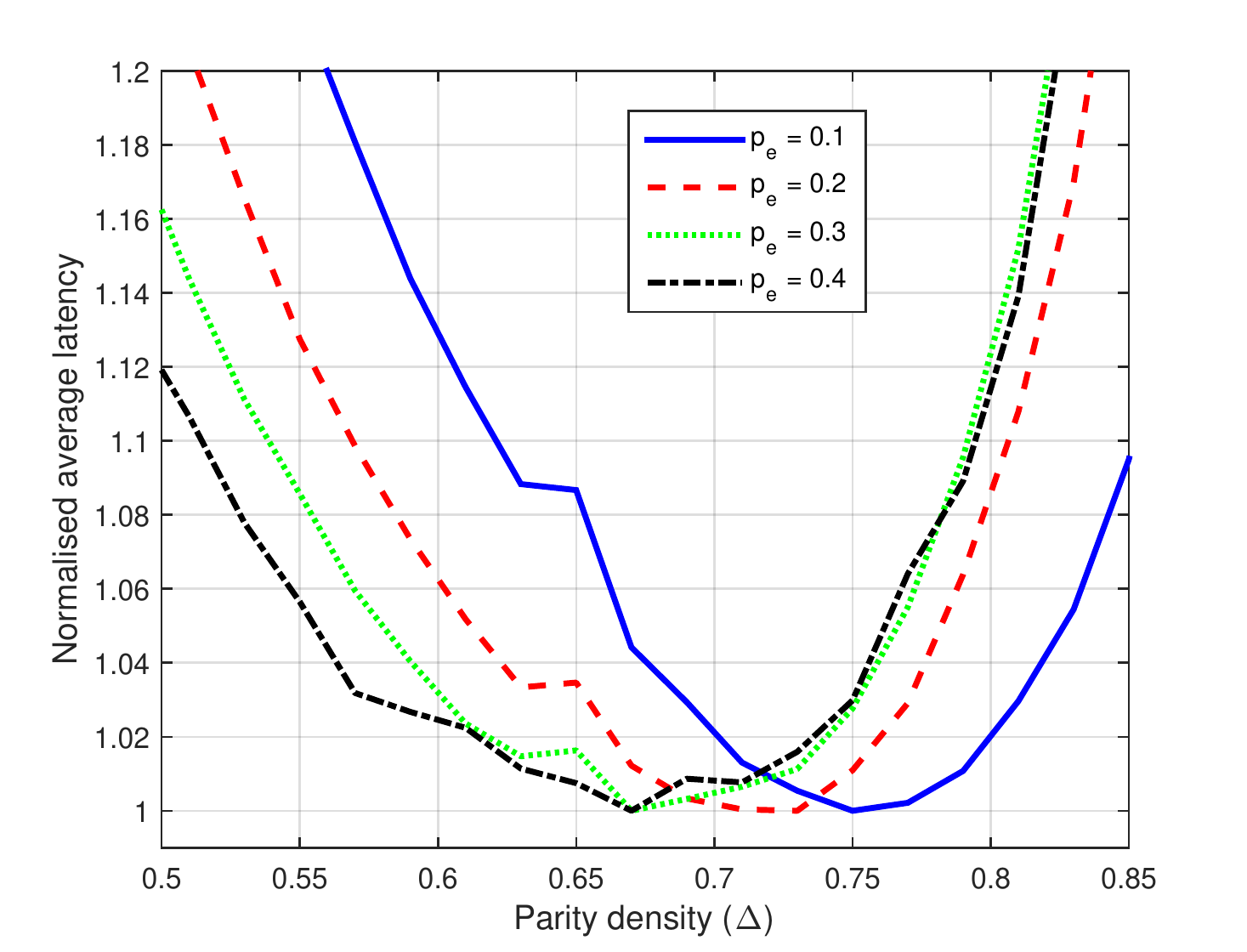}%
\caption{Average normalised latency of recovered data (rate $R=1/2$, memory $W=50$).}%
\label{fig:delta}%
\end{figure}

The most important metric is of course the DRR, as it reflects the number of symbols recovered through the coding. However this does not take into account another important aspect: the latency. While coding can recover lost data symbols, these could be quite old by the time they are decoded. If the data is time-sensitive, the recovery might be unnecessary. While the parity density $\Delta$ does not play a major part for the DRR, it does have an impact on the latency. The range of $\Delta$ for optimum DRR was shown in \cite{Marcelis17} to be quite large; however when adding the latency metric, we can show that there are optimal values. In Figure \ref{fig:delta}, the normalised\footnote{The latency is normalised by its minimal value since the absolute value varies with the packet loss $p_e$, which would make the curves difficult to compare.} average latency is shown as a function of the density $\Delta$. The average latency is only measured over recovered (decoded) data symbols; data symbols that are received over the channel (and hence has latency zero) are not included. The optimal value of $\Delta$ depends on the packet loss rate $p_e$ of the channel but a choice of $\Delta \approx 0.7$ works for most cases. It's worth noting that the latency can increase substantially if the density is chosen too small. 

\subsection{Code rate}

\begin{figure}[thb]%
\centering
\includegraphics[width=\linewidth]{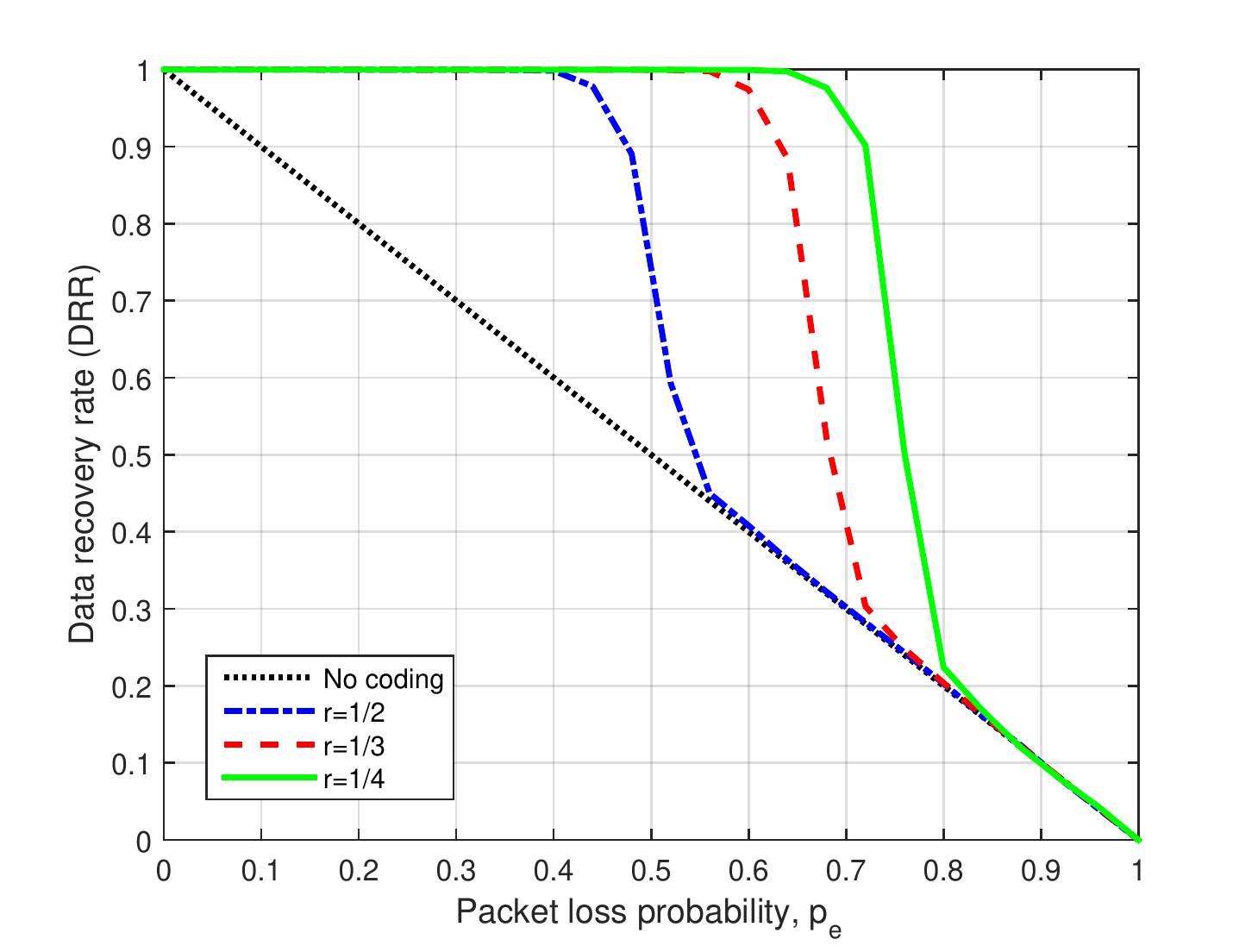}%
\caption{Data recovery rate as a function of packet loss probability for different code rates ($W=50$, $\Delta=0.5$).}%
\label{fig:code_rate}%
\end{figure}

As mentioned earlier, it is possible to use lower code rates than $R=1/2$ to offer more protection. The parity symbols are independently generated with the same memory and density (individual values do not seem to offer any advantages). The DRR for three different code rates are shown in Figure \ref{fig:code_rate}; as expected, the performance improves with lower code rate (more redundancy).

The asymptotic behaviour of the coding scheme can actually be predicted by using existing results from fountain code theory. Full recovery of the missing packets is possible when the generator matrix $\mathbf{G}$ has full rank. In order for the probability of having a full rank random $K \times N$ binary matrix to be at least $1-\delta$, we must have $N \geq K +\log_2 \frac{1}{\delta}$ columns \cite{MacKay05}. This excess amount of packets, $N-K$, makes it increasingly likely that $\mathbf{G}$ has full rank. Assume the above coding scheme with rate $R$ and that $n$ packets have been transmitted, each with $1/R$ symbols. With $p_e$ denoting the packet loss probability, the expected number of received symbols is $(1-p_e)n/R$. Hence we need

\begin{eqnarray}
&& \left( 1 - p_e \right) \frac{n}{R} \geq n + \log_2 \frac{1}{\delta}  \nonumber \\
&& \Rightarrow R \leq \frac{n \left( 1-p_e \right)}{n - \log_2 \delta} \rightarrow 1 - p_e
\label{eq:succ_dec}
\end{eqnarray}
where the last expression is the limit as the number of packets grows. Hence it is clear that as the packet loss probability increases, the code rate must be reduced to maintain the same probability of successful decoding. 

\subsection{System performance effects}

Another aspect of the code rate is how the redundancy affects system performance. Since the payload of the packets is increased by a factor $1/R$, this could actually negatively influence the performance. For small payloads (relative to the overhead such as preamble, headers, etc), it has marginal effects. However if the payload is substantially larger than the overhead, this effectively makes the packet $1/R$ times larger. This in turn will increase the probability of packet collisions since it is more likely that two packets will overlap. 

Consider a simplified model where there are $l$ end-devices and $m$ slots in the time-frequency grid. A packet collision will occur if two or more packets occupy the same slot\footnote{We ignore partial overlaps and relative signal strength. For a more in-depth outage analysis, see \cite{Georgiou17}.}. The probability that we have only one packet in a slot is then

\begin{equation}
\left( 1 - \frac{1}{m} \right) ^{l-1} \rightarrow \exp ^{-(l-1)/m},~ l \gg 1
\end{equation}
and consequently the packet loss probability is
\begin{equation}
p_e \approx 1 - \exp^{-l/m}.
\end{equation}

\begin{figure}[thb]%
\centering
\includegraphics[width=\linewidth]{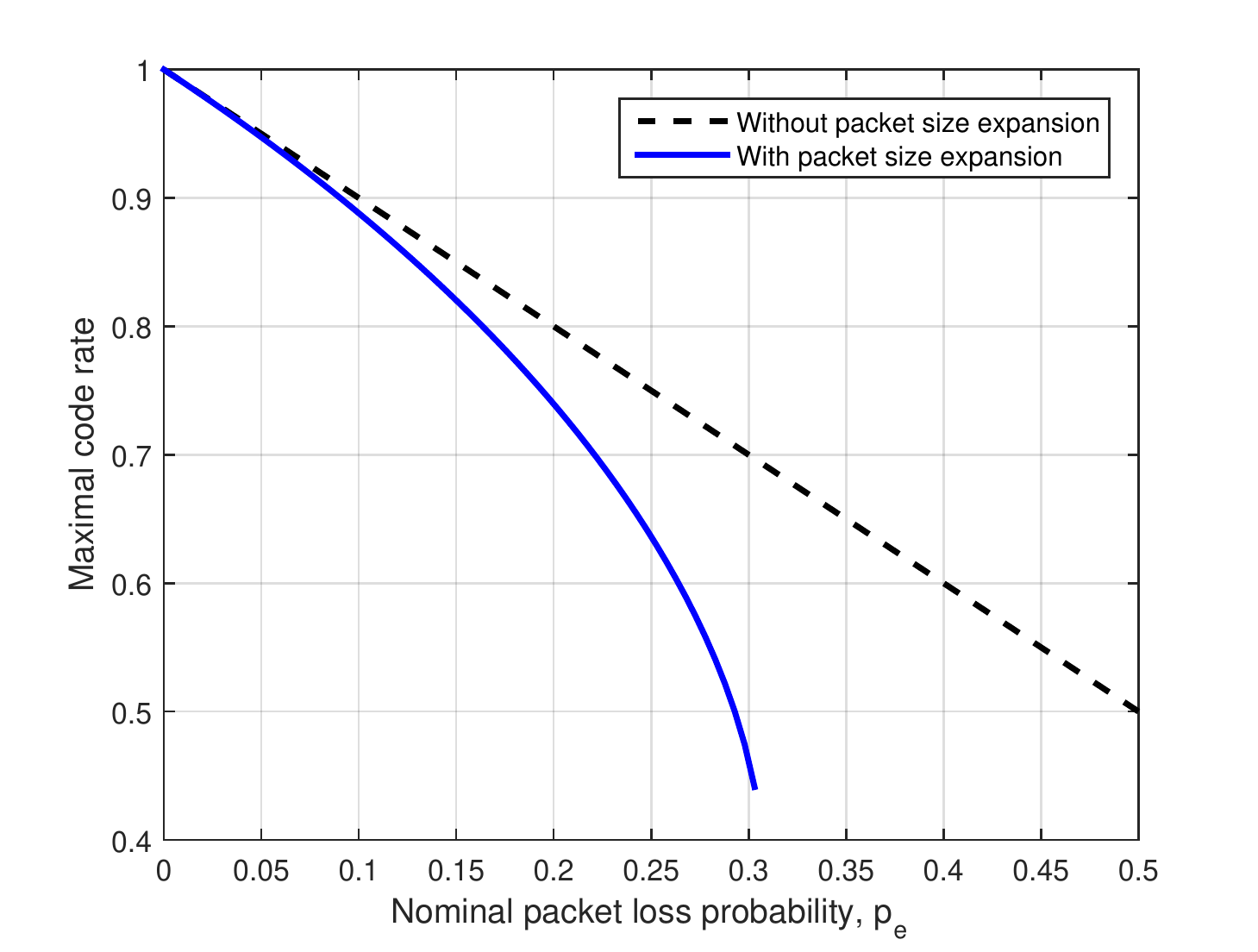}%
\caption{Maximum effective code rate with and without packet size expansion.}%
\label{fig:max_code_rate}%
\end{figure}

By increasing the packet size by $1/R$ times, we effectively reduce the number of slots to $mR$. The packet loss probability now becomes
\begin{equation}
p_e' \approx 1 - \exp^{-l/mR} = 1 - \left( 1-p_e \right)^{1/R}.
\end{equation}
The condition for successful decoding, \eqref{eq:succ_dec}, now becomes
\begin{equation}
R \leq 1 - p'_e(R) = \left( 1-p_e \right)^{1/R}.
\end{equation}
Hence we can relate the nominal packet loss probability $p_e$ to the maximum code rate when packet size expansion is taken into account. This means that the effective maximum code rate is
\begin{equation}
R_{\text{max}} = \arg \max_{R} \left\{ R \left| ~R < \left( 1-p_e \right)^{1/R} \right. \right\}.
\end{equation} 
This is shown in Figure \ref{fig:max_code_rate}. It is worth noting that if the nominal packet loss probability exceeds $p_e > 0.3$, the coding introduces an unrecoverable increase in packet loss and no code rate exists that can successfully (on average) recover lost packets.

\section{Decoding\label{sec:decoding}}

Recovering the lost data symbols is possible if the generator matrix $\mathbf{G}$ has full rank, \emph{i.e.}, it is invertible. In this case, we can solve for the unknown symbols and recover all data symbols. In this section we briefly discuss the techniques and complexity of the different methods at our disposal. It is important to remember that all arithmetic is in $GF(2)$, which can sometimes simplify the job.

\subsection{Optimal decoding}

Solving a system of linear equations can be done by Gaussian elimination \cite{Strang88}. This is based on the principle that columns of $\mathbf{G}$ can be added to each other to create a triangular structure (top left half of the matrix contains only zeroes) of new matrix. Once this is achieved, backsubstitution is used to solve the equations. However sometimes it is not possible to achieve a triangular structure, in which pivoting (swapping columns/rows to obtain a nonzero value on the diagonal) is necessary. For an $n \times n$ matrix, a straightforward implemention of Gaussian elimination has complexity $\mathcal{O} \left( n^3 \right)$. However with optimised parallelised hardware, this can be brought down to $\mathcal{O} \left( n^2 \right)$ \cite{Bogdanov06}. 

For the systems described in this paper, it is important to note that due to the low duty cycle of the end-devices, there might be quite some time between packets. This can be used to process as much of the decoding matrix as possible, which alleviates the need for excessive processing when new parity symbols arrive. If fewer parity symbols than lost data symbols are available, decoding is not possible and the decoding matrix must be kept until new redundancy is obtained. By using elementary row and column operations \cite{Strang88}, the decoding matrix $\mathbf{G}'$ of size $s$ (missing data symbols) by $t$ (linearly independent parity equations) can be arranged as

\begin{equation}
\mathbf{G}' = \left( 
\begin{array}{c}
\mathbf{I}_t \\
\mathbf{A}
\end{array}
\right)
\end{equation}
where $\mathbf{I}_t$ is an $t \times t$ identity matrix and $\mathbf{A}$ is a $(s-t) \times t$ matrix. When a new parity equation is available through a received parity symbol, $\mathbf{G}'$ can easily be updated as
\begin{equation}
\mathbf{G}'' = \left( 
\begin{array}{c}
\mathbf{I}_t \\
\mathbf{A}
\end{array}
\mathbf{g}
\right)
\rightarrow
\left( 
\begin{array}{c}
\mathbf{I}_{t+1} \\
\mathbf{A'}
\end{array}
\right).
\end{equation}
If the new vector $\mathbf{g}$ does not allow such a transformation, it is linearly dependent on the columns in $\mathbf{G}'$ and can be discarded. Decoding of symbols is now possible if any column of $\mathbf{G}''$ has only one nonzero element; this means that this equation has only one variable and hence it can be solved for. The trivial case is of course when the lower matrix $\mathbf{A}'$ is empty or all-zero and all $t+1$ variables can be solved. The rows and columns of $\mathbf{G}''$ corresponding to the solved variables and used equations, respectively, can be removed and the remaining matrix is kept for future decoding. Note that this 'continuous` Gaussian elimination avoids duplicate operations and simplifies finding linearly dependent equations.  

In the next section, we will discuss other methods to reduce complexity as well as a novel approach to the decoding problem in this paper.

\subsection{Reduced complexity decoding}
\label{sec:red_comp_dec}

Fountain codes were initially designed to have linear encoding and decoding time \cite{MacKay05}. This was achieved by using LT decoder, which is a type of {\it message passing} algorithm. Consider the following linear system of equations in $GF(2)$ \cite{MacKay05}
\begin{equation}
\left\{
\begin{array}{rcl}
x_1 &=& 1  \\
x_1 \bigoplus x_2 \bigoplus x_3 &=& 0  \\ 
x_2 \bigoplus x_3 &=& 1  \\ 
x_1 \bigoplus x_2 &=& 1.  
\end{array}
\right.
\end{equation}
Since the first equation only has one unknown we solve this ($x_1=1$) and replace the variable with this value in the remaining equations.
\begin{equation}
\left\{
\begin{array}{rcl}
x_2 \bigoplus x_3 &=& 1  \\ 
x_2 \bigoplus x_3 &=& 1  \\ 
x_2 &=& 0.  
\end{array}
\right.
\end{equation}
The last equation only has one variable, so solving this ($x_2=0$) and substituting it in the other equations, gives us
\begin{equation}
\left\{
\begin{array}{rcl}
x_3 &=& 1 \\ 
x_3 &=& 1  
\end{array}
\right.
\end{equation}
The remaining equations now only has one variable, so we get $x_3=1$. Solving this system can clearly be done in linear time. 

However there is no guarantee that there will always be an equation with only one variable. If this does not happen, the decoding halts and no further decoding can take place (even if the system has a solution). Early work on fountain codes concerned designing the distribution of chosen data symbols to form the parity symbols (nonzero values in a column of $\mathbf{G}$); this can be done to minimise the chances of the algorithm halting.

Another approach is that if the LT decoder halts, a particular variable can be solved. Wiedemann \cite{Wiedemann86} designed a reduced complexity method for solving one (but not all) variables in a finite field. This was applied in \cite{Lu13} to the LT decoder to restart the algorithm whenever there were no more variables to solve directly. Although this increases the complexity, it was shown in \cite{Lu13} that it does not need to be applied very often with a carefully designed code. 

A further modification of the message passing decoder is the \emph{inactivation} method \cite{Shokrollahi06}. Instead of solving for one of the variables when LT decoder halts, it is labelled inactive and assumed known. The decoder then continues until it halts again, when another variable is labelled inactive and assumed known. Eventually all active variables are solved or rather, they are functions of the $l$ inactive variables. These can be solved with Gaussian elimination (or similar methods) and backsubstituted into the other variables. The advantage is that a much smaller system needs to be solved of size $l \times l$, which can offer huge complexity reductions if $l \ll n$.

\subsection{Suboptimal decoding}  
\label{sec:truncation}
The methods described above all find the solution, \emph{i.e.}, they are optimal in terms of performance. It is also possible to design suboptimal decoders which trade off performance for reduced complexity. The generator matrix is built up as more parity symbols are received. However if some of the lost data symbols are very old, they could be discarded to reduce the size of the matrix to be inverted. This is done by removing the corresponding row and any columns with a one in this row. The latter step is necessary since we do not want parity symbols that depend on the discarded data symbol.

\begin{figure}[thb]%
\centering
\includegraphics[width=\linewidth]{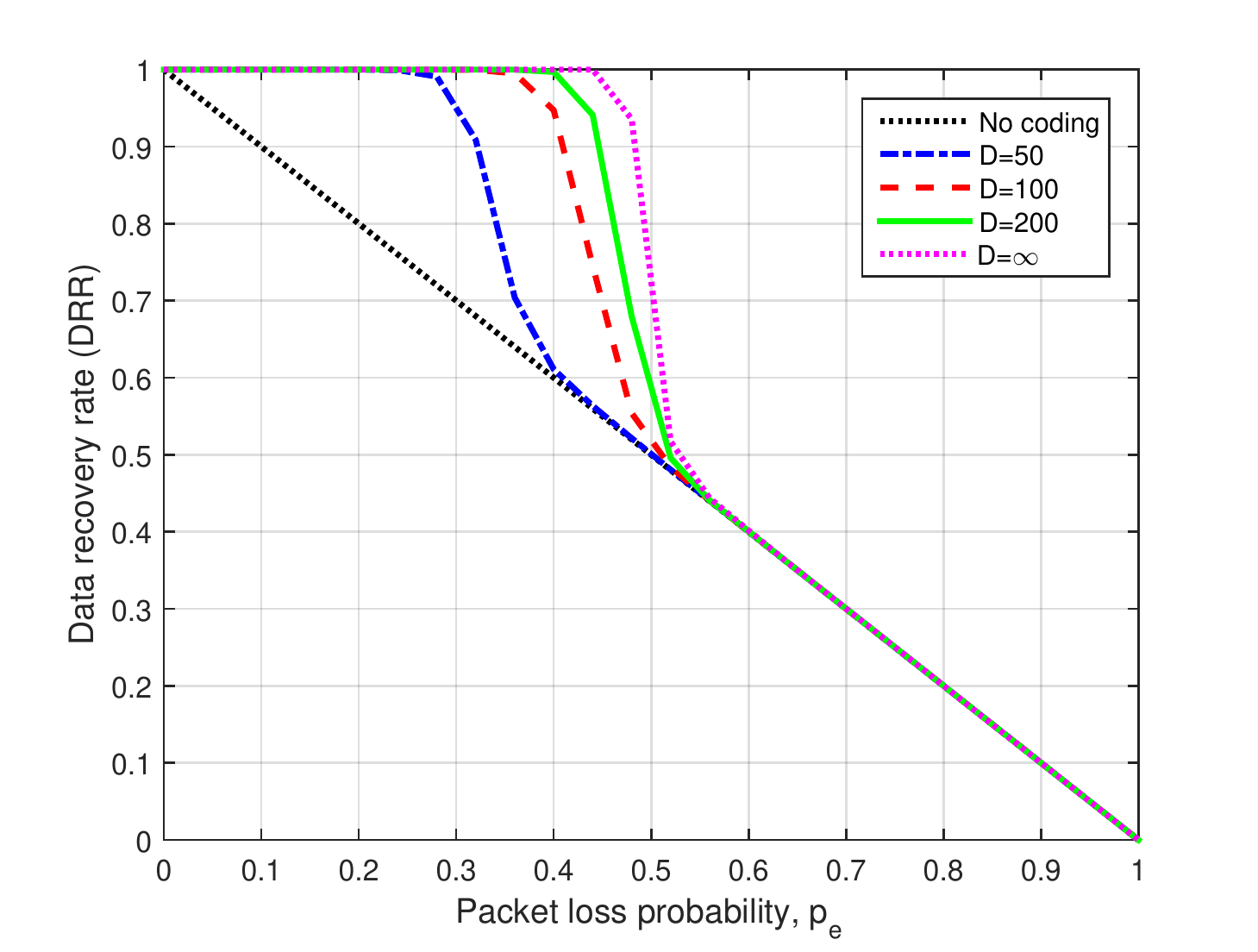}%
\caption{DRR performance when a maximum decoding delay is introduced (rate $R=1/2$, memory $W=50$).}%
\label{fig:suboptimal_drr}%
\end{figure}

This novel complexity-reducing technique also has a practical side effect. If the end-devices have time-sensitive data to send, it will not make sense to wait until a sufficient number of parity symbols have been received so decoding can start. Instead these symbols can be discarded if they are too old as they are of little value. The downside is of course that the data recovery rate is reduced, which is evident from Figure \ref{fig:suboptimal_drr}. 

\begin{figure}[thb]%
\centering
\includegraphics[width=\linewidth]{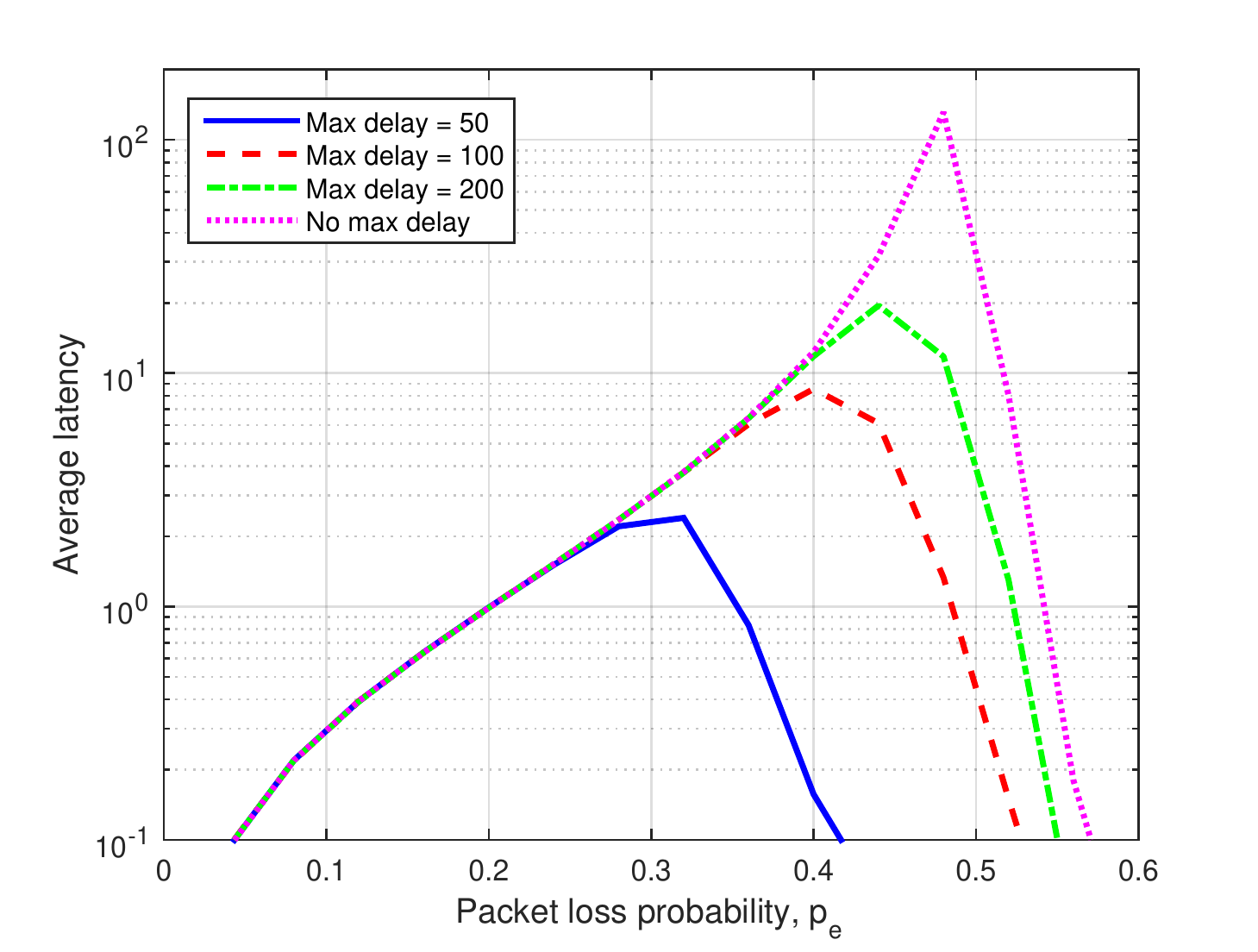}%
\caption{Average latency of recovered data (rate $R=1/2$, memory $W=50$).}%
\label{fig:average_latency}%
\end{figure}

The decoding delay is the age of a recovered data symbol in the decoding buffer; if symbol $n-d$ is decoded at time $n$, the delay is $d$. Note that this is measured in terms of packets. If the packets are sent on a regular basis, the actual delay is simply the $d$ times the packet transmission interval. Otherwise it must be measured using, \emph{e.g.}, time stamps. If a symbol in the decoding buffer has a delay that exceeds a predetermined value (it is too old to be valuable even if it's recovered), it is discarded and the decoding matrix is pruned as described above. Despite the loss in DRR, it will have benefits in terms of latency and complexity. This is shown in Figure \ref{fig:average_latency}, where the average latency can be limited by truncating the decoding process. As the maximum allowed delay is increased, the average latency is also increased. It is worth noting that the average latency drops as the packet loss increases; this is simply due to the fact that more symbols can not be decoded and hence do not contribute towards the latency. For very high packet losses (above the code rate), the latency goes down to zero since no lost symbols can be recovered (only the ones received over the channel are available).

\begin{figure}[thb]%
\centering
\includegraphics[width=\linewidth]{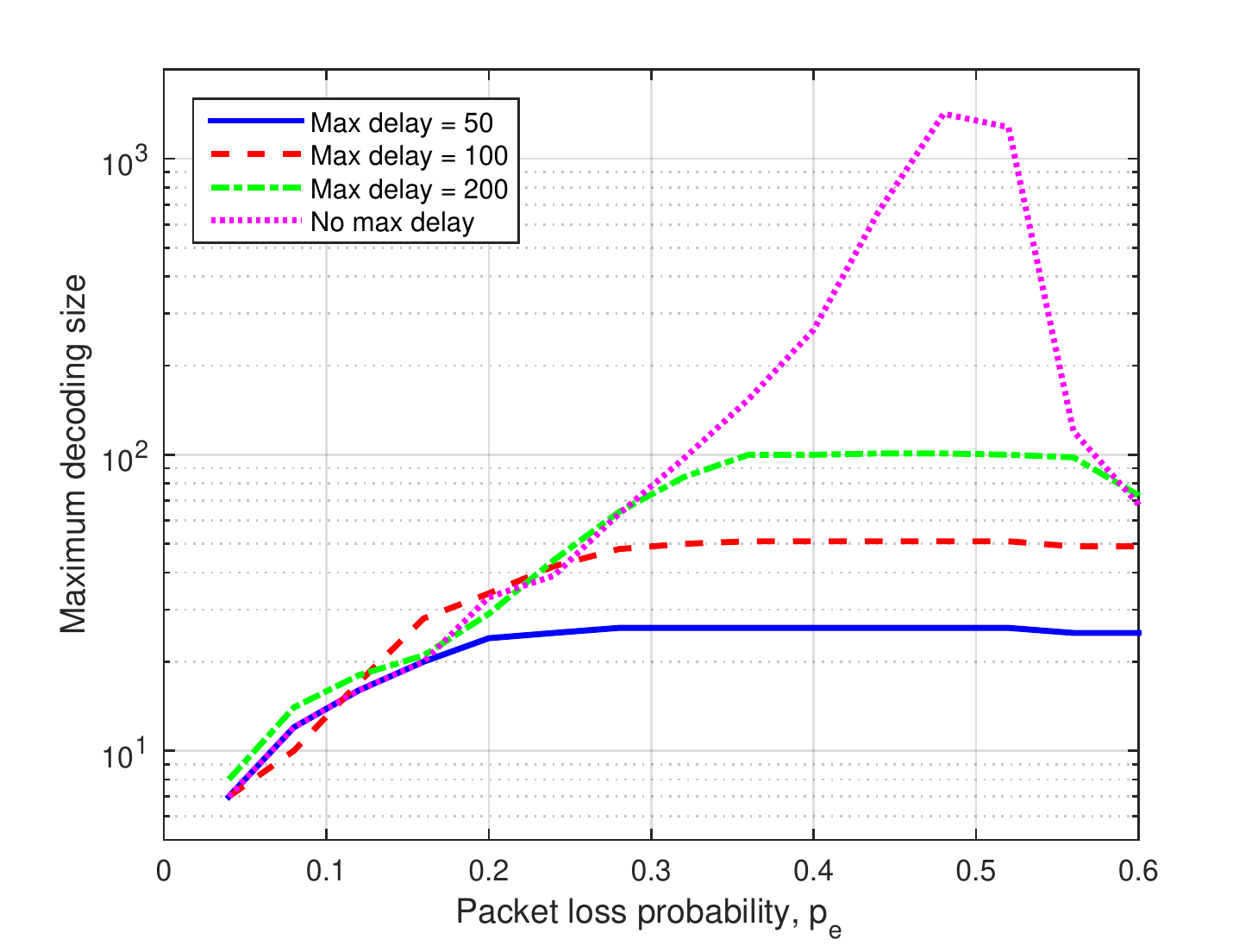}%
\caption{Maximum size of the generator matrix for decoding (rate $R=1/2$, memory $W=50$).}%
\label{fig:max_decoding}%
\end{figure}

This has also a beneficial impact on the complexity, which is shown in Figure \ref{fig:max_decoding}; the complexity here is defined as the average size of the decoding buffer (number of symbols yet to decode). Regardless of decoder choice, this is an indication of the computational (and memory) burden of the data recovery. As can be seen, the novel scheme reduces the complexity as it limits the size of the decoding problem; this might have significant advantages when it comes to implementation.

%
%

\begin{table}
		\caption{Comparison of different decoders.  \label{tab:decoders}}
	\centering
		\begin{tabular}{|l||l|l|l|} \hline
			\bf{Decoder} & \bf{Performance} & \bf{Complexity} & \bf{Latency} \\ \hline \hline 
			Gaussian elimination & Optimal & High & Medium \\ \hline
			LT-W \cite{Lu13} & Optimal & Medium & Medium \\ \hline
			Inactivation \cite{Shokrollahi06} & Optimal & Medium & Medium \\ \hline
			Message passing \cite{Luby02} & Suboptimal & Low & High \\ \hline
			Truncation (Section \ref{sec:truncation}) & Suboptimal & Medium & Low \\ \hline 
		\end{tabular}
\end{table}

The different decoders and their properties are compared in Table \ref{tab:decoders}, where we have listed their relative performance, complexity and latency.

\section{Conclusions\label{sec:conclusions}}
In this paper we have analysed the application layer coding scheme for low power wide area networks introduced in \mbox{\cite{Marcelis17}}. We have extended their study to include latency and the effects of decreased code rates as well as decoder complexity. The latency was shown to increase exponentially with the packet loss rate but a novel decoding scheme can reduce this with a small loss in data recovery. This new scheme can also limit the decoding complexity and memory requirements; a quantative comparison between different decoding options was also given. We also showed that increased packet loss can not be solely combated by introducing more redundancy; at some point the increased packet size will cause an irreparable number of packet collisions.

\section{Acknowledgements}

The authors would like to acknowledge the fruitful discussions with their
colleagues at Toshiba Research Europe and the support of its directors. 

\bibliographystyle{IEEEtran}
\bibliography{IEEEabrv,iotfc}

\end{document}